\documentstyle[12pt]{article}
\catcode`@=11
\newcount\@tempcntc
\def\@citex[#1]#2{\if@filesw\immediate\write\@auxout{\string\citation{#2}}\fi
  \@tempcnta\z@\@tempcntb\m@ne\def\@citea{}\@cite{\@for\@citeb:=#2\do
    {\@ifundefined
       {b@\@citeb}{\@citeo\@tempcntb\m@ne\@citea\def\@citea{,}{\bf ?}\@warning
       {Citation `\@citeb' on page \thepage \space undefined}}%
    {\setbox\z@\hbox{\global\@tempcntc0\csname b@\@citeb\endcsname\relax}%
     \ifnum\@tempcntc=\z@ \@citeo\@tempcntb\m@ne
       \@citea\def\@citea{,}\hbox{\csname b@\@citeb\endcsname}%
     \else
      \advance\@tempcntb\@ne
      \ifnum\@tempcntb=\@tempcntc
      \else\advance\@tempcntb\m@ne\@citeo
      \@tempcnta\@tempcntc\@tempcntb\@tempcntc\fi\fi}}\@citeo}{#1}}
\def\@citeo{\ifnum\@tempcnta>\@tempcntb\else\@citea\def\@citea{,}%
  \ifnum\@tempcnta=\@tempcntb\the\@tempcnta\else
   {\advance\@tempcnta\@ne\ifnum\@tempcnta=\@tempcntb \else \def\@citea{--}\fi
    \advance\@tempcnta\m@ne\the\@tempcnta\@citea\the\@tempcntb}\fi\fi}
\catcode`@=12

\begin{document}

\begin{flushright}
RAL-TR/96--021\\
March 1996
\end{flushright}

\begin{center}
{\Large{\bf  Resonant CP-Violating Scalar--Pseudoscalar}}\\[0.4cm]
{\Large{\bf   Transitions at {\boldmath $\mu^+\mu^-$} Colliders }}\\[1.5cm]
{\large Apostolos Pilaftsis}\footnote[1]{E-mail address: 
pilaftsis@v2.rl.ac.uk}\\[0.4cm]
{\em Rutherford Appleton Laboratory, Chilton, Didcot, Oxon, OX11 0QX, UK}
\end{center}
\vskip0.6cm
\centerline {\bf ABSTRACT}
A $\mu^+\mu^-$ collider is an appealing machine to probe resonant CP-violating
transitions of a CP-even Higgs particle into the $Z$ boson or into another
CP-odd Higgs scalar. These phenomena are studied within a manifestly
gauge-invariant approach implemented by the pinch technique. The CP
invariance of an extended Higgs sector motivated by supersymmetric E$_6$
models is assumed to be broken radiatively by the presence of heavy Majorana
fermions. CP violation originating from Higgs-$Z$ mixing is found to be very
modest, whereas CP-number violating transitions involving Higgs scalars only
can be resonantly enhanced up to order of unity.\\[0.4cm] 

\newpage

Recently, much research and theoretical effort have been put into the design and
the physics capabilities of a muon collider, which could potentially serve as
an important Higgs factory \cite{mumu}. Given the technical facility of a
variable centre-of-mass (c.m.) energy for such a machine, it has been argued
\cite{mumu} that one could exploit the resonant enhancement of an $s$-channel
interaction to copiously produce Standard Model (SM) Higgs-bosons, $H$, in the
mass range $100\le M_H\le 200$ GeV, and/or explore the existence of non-SM Higgs
bosons with $0.2\le M_H\le 1$ TeV, which couple to $\mu$ significantly. The 
most compelling extensions of the SM with naturally large $H\mu\mu$ couplings
are those having an underlying supersymmetric (SUSY) origin. In order to
ensure the absence of the triangle anomalies and give mass to both up and down
quarks, the Higgs sector of a SUSY scenario must contain at least two Higgs
doublets, giving rise to CP-even and CP-odd Higgs scalars. Therefore, a
muon collider may be the most ideal place to search for large Higgs-$Z$-boson
mixing effects and, more interestingly, observe transitions between Higgs
scalars with opposite CP quantum numbers. In fact, if quantum effects allow
for a CP-even Higgs scalar, $H$, to go into the CP-odd $Z$ boson or another
CP-odd scalar, $A$ say, such a transition alone would signify CP/T violation
in a CPT-invariant theory \cite{Kabir}. In the SM, there is no $HZ$ mixing up
to two-loop electroweak order. The reason is that a non-trivial CP-odd
rephasing invariant combination of Cabbibo-Kobayashi-Maskawa matrix
elements is required inside the fermionic loops. However, in natural
extensions of the SM, involving Majorana fermions \cite{IKP} or more than one
Higgs doublet \cite{CK}, a $HZ$-mixing effect may occur in the decays of the
$H$ into top-quark pairs. 

In this paper, we study the possibility of CP-violating $HZ$ and/or $HA$
transitions in a model, in which the CP invariance of the Higgs sector is
broken radiatively by the presence of heavy Majorana fermions. Models of the
kind are the minimal SUSY SM (MSSM), in which Majorana fermions may be
identified with the heavy neutralinos, or other scenarios inspired by
SUSY-E$_6$ theories, which predict heavy Majorana neutrinos \cite{E6,RM}
at the TeV mass scale. Here, we will work on the latter realization
\cite{E6Rb}. To be specific, we adopt the CP-violating scenario of \cite{IKP}
for the neutrino sector, which may resemble the model discussed by the authors
in Ref.\ \cite{RM} at the electroweak scale. The model contains three heavy
Majorana neutrinos, denoted here by $N_1$, $N_2$, and $N_3$, from which $N_1$
is predominantly a sequential isodoublet, whereas $N_2$ and $N_3$ are mainly
singlets under SU(2)$_L$. Furthermore, we consider that the Lagrangians
governing the interactions between $N_i$ (with $i=1,2,3$) and $H$, $A$, and
the would-be Goldstone boson $G^0$, have the following generic form
\cite{zpc}: 
\begin{eqnarray}
\label{ANN}
{\cal L}_{A} &=&  \frac{i g}{4M_W}\ A\, \chi^u_A\, 
\sum_{i,j=1}^{3}
\bar{N}_i \Big[ \gamma_5\, (m_i+m_j)\Re e C_{ij}
+\ i(m_j-m_i)\Im m C_{ij} \Big] N_j\ ,\\
\label{HNN}
{\cal L}_H &=& -\ \frac{g}{4M_W}\ H\, \chi^u_H\, 
\sum_{i,j=1}^{3}
\bar{N}_i \Big[ (m_i+m_j)\Re e C_{ij}
+\ i\gamma_5 (m_j-m_i)\Im m C_{ij} \Big] N_j\ ,
\end{eqnarray} 
where the parameters $\chi^u_{A,H}$ are related with the vacuum expectation
values (VEVs) of the Higgs fields in an extended Higgs sector and $C_{ij}$ are
mixing matrices defined in \cite{zpc}. The coupling $G^0N_iN_j$ can be
recovered from Eq.\ (\ref{ANN}), if we set $\chi^u_H=1$ and $A\equiv G^0$.
This model has a non-trivial CP-violating phase contained in the rephasing
invariant quantity 
           $\Im m C^2_{N_1N_2}=\sin\delta_{CP} |C_{N_1N_2}|^2$ \cite{IKP}, 
which is taken to be maximum of order one.

To analyze CP violation originating from $HZ$ and/or $HA$ mixing, we have to
find an observable sensitive to these effects. Assuming that the facility of
having longitudinally polarized muon beams will be available without much loss
of luminosity, we can define the CP asymmetry 
\begin{equation}
\label{CPobs}
{\cal A}_{CP}\ =\ \frac{\sigma (\mu^-_L\mu^+_L\to f\bar{f})\ -\
\sigma (\mu^-_R\mu^+_R\to f\bar{f})}{\sigma (\mu^-_L\mu^+_L\to f\bar{f})\ +\
\sigma (\mu^-_R\mu^+_R\to f\bar{f})}\, .
\end{equation}
We must emphasize that ${\cal A}_{CP}$ is a genuine observable of CP violation
if one is able to tag on the final fermion pair $f\bar{f}$ ({\em e.g.},
$\tau^+\tau^-$, $b\bar{b}$, or $t\bar{t}$), since the helicity states
$\mu^-_L\mu^+_L$ transform into $\mu^-_R\mu^+_R$ under CP in the
c.m.\ system. Similar CP/T-violating observables based on T-odd
aplanarities at $e^+e^-$ machines were considered by the authors of Ref.\
\cite{Gavela}, who suggested to look for CP violation in vector and
axial-vector currents. Here, we require, however, that both muons are
left-handed or right-handed polarized. Similar ideas have been applied to
study CP violation in the top-pair production at LHC and TeV-$e^+e^-$
colliders \cite{Peskin,CK,IKP} and, more recently, to muon colliders
\cite{CPmumu} as well. 

In our analysis, we consider a manifestly gauge-invariant approach for
resonant transitions~\cite{JP&AP}, which is implemented by the pinch technique
(PT) \cite{JMC}. This approach is free from CP-odd gauge artifacts; it
reassures the absence of a $HZ$ and/or $HA$ mixing in a CP-invariant and
anomaly-free theory, thus preserving the discrete symmetries of the classical
action after quantization. Furthermore, we make use of a mechanism for
resonant CP violation induced by particle widths in scattering processes,
which was discussed in \cite{AP} in connection with top-quark production
and decay some time ago \cite{AS}. This mechanism of CP violation
gives rise to a resonant enhancement for certain CP-violating observables, such
as ${\cal A}_{CP}$ in Eq.\ (\ref{CPobs}), and so yields measurable effects for
a wide range of heavy Higgs masses as we will see below. 

We now discuss in short how resummation involving $ZH$ mixing takes place
within our approach \cite{JP&AP}. There are PT identities that can be employed 
to convert $ZH$ and $ZZ$ strings into $G^0H$ and $G^0G^0$ ones \cite{PS} before
resummation occurs. These identities are 
\begin{equation}
p^\mu\widehat{\Pi}_\mu^{ZH}+i M^0_Z\widehat{\Pi}^{G^0H} = 0\, ,\quad
p^\mu p^\nu\widehat{\Pi}_{\mu\nu}^{ZZ}-(M^0_Z)^2\widehat{\Pi}^{G^0G^0} = 0\, ,
\mbox{and}\quad  
p^\mu \Gamma^{Zf\bar{f}}_\mu = -iM^0_Z \Gamma^{G^0f\bar{f}}\, ,
\end{equation}
where $p^\mu$ of the $Z$ boson always flows into the fermionic vertex and
$M^0_Z$ is the bare $Z$-boson mass. If
$\widehat{\Pi}_\mu^{ZH}(p)=p_\mu\widehat{\Pi}^{ZH}(p^2)$ and
$\widehat{\Pi}^{ZZ}_{\mu\nu}(p)=t_{\mu\nu}(p)\widehat{\Pi}^{ZZ}_T(p^2)
+\ell_{\mu\nu}(p)\widehat{\Pi}^{ZZ}_L$, with $t_{\mu\nu}(p)=-g_{\mu\nu} +p_\mu
p_\nu/p^2$ and $\ell(p)=p_\mu p_\nu/p^2$, then
$\widehat{\Pi}^{ZH}(p^2)=-iM^0_Z\widehat{\Pi}^{G^0H}(p^2)/p^2$ and
$\widehat{\Pi}^{ZZ}_L(p^2)=(M^0_Z)^2\widehat{\Pi}^{G^0G^0}(p^2)/p^2$. In this
context, it worth stressing the fact that $p^\mu\widehat{\Pi}^{\gamma G^0}_\mu
(p)= p^\mu\widehat{\Pi}^{\gamma H}_\mu (p)= 0$, within the PT framework, which
implies the absence of $\gamma H$ and $\gamma G^0$ mixing, {\em i.e.},
$\widehat{\Pi}^{\gamma H}_{\mu} = \widehat{\Pi}^{\gamma G^0}_\mu =0$,
independently of whether CP-violating interactions are present in the theory.
As a result, one is left with solving the simple coupled Dyson-Schwinger
equation system, in which only $H$ and $G^0$ mix, {\em i.e.}, 
\begin{equation}
\label{Delta}
[\hat{\Delta} (p^2) ]^{-1}\ =\ \left[
\begin{array}{cc}
p^2+\widehat{\Pi}^{G^0G^0}(p^2) & \widehat{\Pi}^{G^0H}(p^2)\\
\widehat{\Pi}^{HG^0}(p^2) & p^2-(M^0_H)^2+\widehat{\Pi}^{HH}(p^2)
\end{array} \right]\, .
\end{equation}
Inverting this matrix, we find
\begin{eqnarray}
\label{PTprop}
\hat{\Delta}_{G^0}(p^2) &=& \Big\{\, p^2+\widehat{\Pi}^{G^0G^0}(p^2)-
[\widehat{\Pi}^{G^0H}(p^2)]^2/[p^2-(M^0_H)^2+\widehat{\Pi}^{HH}(p^2)]
\Big\}^{-1}\, ,\nonumber\\
\hat{\Delta}_{H}(p^2) &=& \Big\{\, p^2-(M^0_H)^2+\widehat{\Pi}^{HH}(p^2)-
[\widehat{\Pi}^{G^0H}(p^2)]^2/[p^2+\widehat{\Pi}^{G^0G^0}(p^2)]
\Big\}^{-1}\, ,\nonumber\\
\hat{\Delta}_{G^0H}(p^2) &=&-\widehat{\Pi}^{G^0H}(p^2) 
\Big\{ [p^2-(M^0_H)^2+\widehat{\Pi}^{HH}(p^2)]
[p^2+\widehat{\Pi}^{G^0G^0}(p^2)]\nonumber\\
&&-[\widehat{\Pi}^{G^0H}(p^2)]^2\, \Big\}^{-1}\, .
\end{eqnarray}
The above considerations can be extended to include additional Higgs scalars,
such as the physical CP-odd scalar, $A$, and so describe $HA$-mixing effects. 
In such a case, the inverse propagator in Eq.\ (\ref{Delta}) becomes a 
$3\times 3$ matrix. 

Within the PT, it is known \cite{BFM} that the analytic expression of $\Im
m\widehat{\Pi}^{HH}(s)$ coincides, to one loop, with that of the background
field gauge for $\xi_Q=1$. In this way, we obtain for the different
channels 
\begin{eqnarray}
\label{Piff}
\Im m\widehat{\Pi}^{HH}_{(f\bar{f})}(s) &=& \frac{\alpha_w N^f_c}{8} 
(\chi^f_H)^2\, s\, \frac{m^2_f}{M^2_W} \left(1-\frac{4m^2_f}{s}\right)^{3/2}
\theta (s-4m^2_f )\, ,\\
\label{PiVV}
\Im m\widehat{\Pi}^{HH}_{(VV)}(s) &=& \frac{n_V\alpha_w}{32}(\chi^V_H)^2
\frac{M^4_H}{M^2_W} \left(1-\frac{4M^2_V}{s}\right)^{1/2}\nonumber\\
&&\times \left[ 1+4\frac{M^2_V}{M^2_H}-4\frac{M^2_V}{M^4_H}(2s-3M^2_V)
\right] \theta (s-4M^2_V )\, ,
\end{eqnarray}
where $\alpha_w=g^2/4\pi$, $n_V=2$, 1 for $V\equiv W$, $Z$, respectively, and
$N^f_c=1$ for leptons and 3 for quarks. In Eqs.\ (\ref{Piff}) and
(\ref{PiVV}), we have parametrized fermionic and bosonic channels by the
model-dependent factors $\chi^f_H$ and $\chi^{W,Z}_H$. By analogy, calculation
of $\Im m \widehat{\Pi}^{AA}(s)$ gives $\chi^V_A=0$, while $\chi^f_H$ should
be replaced by $(1-4m^2_f/s)^{-1/2}\chi^f_A$ in Eq.\ (\ref{Piff}). There
may also be other channels involving the $HZA$ vertex, which are, however,
considered to be phase-space suppressed in the kinematic region $M_H\simeq
M_A$, relevant for resonant CP violation. The complete treatment including
these effects as well as other SUSY refinements will be given elsewhere 
\cite{SUSYAP}.

Considering the Lagrangians (\ref{ANN}) and (\ref{HNN}), it is straightforward
to calculate the CP-violating $HG^0$ and/or $HA$ mixing in our model, viz. 
\begin{equation}
\label{HZmix}
\frac{\widehat{\Pi}^{AH}(s)}{s}\ =\ -\frac{\alpha_w}{4\pi}\, \chi^u_A\chi^u_H\,
\sum\limits_{j>i}^{3} \Im m C^2_{N_iN_j}\, \sqrt{\lambda_i\lambda_j}\,
\Big[ B_0(s/M^2_W,\lambda_i,\lambda_j)\, +\,
2B_1(s/M^2_W,\lambda_i,\lambda_j)\Big]\, ,
\end{equation}
where $\hat{\Pi}^{AH}(s)=\hat{\Pi}^{HA}(s)$, $\lambda_i=m^2_{N_i}/M^2_W$, and
$B_0$ and $B_1$ are the usual Veltman-Passarino loop functions, expressed in
the convention of Ref.\ \cite{BAK}. Again, the $HG^0$ transition is recovered
from Eq.\ (\ref{HZmix}) by setting $\chi^u_A=1$. 

From Fig.\ 1, it is not difficult to see that in the case of $HZ$ mixing, only
two diagrams can contribute constructively to ${\cal A}_{CP}$ through the
interference of the $G^0$-exchange graph with the amplitude depending on
$\hat{\Delta}_{HG^0}$. The reason is that contraction of a scalar current with
a pseudoscalar one vanishes identically. Substituting 
Eqs.\ (\ref{PTprop})--(\ref{HZmix}) into Eq.\ (\ref{CPobs}) yields 
\begin{equation}
\label{ACP}
{\cal A}_{CP}(s)\ =\ -2\, \frac{\widehat{\Pi}^{G^0H}(s)}{s}\ 
\frac{\Im m\widehat{\Pi}^{HH}(s)}{s}\, .
\end{equation}
From Eq.\ (\ref{ACP}), we find that CP asymmetries can be large only for heavy
Higgs boson masses far above the two-real $W$-boson production threshold,
since the Higgs width is then comparable to $M_H$. To give an example, we find
${\cal A}_{CP}\simeq 2.\ 10^{-2}$ for $M_H=500$ GeV and $m_{N_1,N_2,N_3}=0.5,\
1.5,\ 3$ TeV, while the production cross-section is $\sigma \simeq 1$ fb. It
seems that one is unlikely to observe $HZ$-mixing effects, even if assuming a
high integrated luminosity of 50 fb$^{-1}$ for the muon collider \cite{mumu}. 

The situation changes drastically if the heaviest CP-even $H$ mixes
with a CP-odd Higgs scalar, $A$, especially when $M_A>2M_Z$. The
latter is very typical within SUSY unified theories \cite{SUSYGUT}. In
our minimally extended SUSY model, $H$ will couple predominantly to
fermions, and naturally be degenerate with $A$, {\em i.e.}\ $M_H\simeq
M_A$ \cite{JFG}. Moreover, the coupling parameters will obey the MSSM
relation $\chi^d_H = \chi^d_A = 1/\chi^u_H = 1/\chi^u_A = \tan\beta$.
Taking these into account, we find that ${\cal A}_{CP}$ behaves at
$s\simeq M^2_H$ as
\begin{equation}
{\cal A}_{CP}\ \sim\ -\, \frac{ 2\widehat{\Pi}^{AH}(s)\,
[ \Im m\widehat{\Pi}^{HH}(M^2_H) - \Im m\widehat{\Pi}^{AA}(M^2_H) ] }{
(M^2_H-M^2_A)^2 + [\Im m\widehat{\Pi}^{AA}(M^2_H)]^2 +
[\Im m\widehat{\Pi}^{HH}(M^2_H)]^2}\, .
\end{equation}
In Fig.\ 2, we have presented cross sections (solid lines) and CP
asymmetries (dotted lines) as a function of the c.m.\ energy,
$\sqrt{s}$, in two different scenarios. We also assume that tuning the
collider energy to the mass of $H$ is feasible, {\em i.e.},
$\sqrt{s}=M_H$. In our estimates, we take $\tan\beta=2$,
$m_{N_1,N_2,N_3}=0.5,\ 1,\ 1.5$ TeV, and $m_t=170$ GeV. We analyze two
reactions: (a) $\mu^+_L\mu^-_L\to b\bar{b}$, for $M_A=170$ GeV and
$\chi^V_H=1$, and (b) $\mu^+_L\mu^-_L\to t\bar{t}$, for $M_A=400$ GeV
and $(\chi^V_H)^2=0.1$. In reaction (a), one can have a significant
CP-violating signal if $M_H=170\pm 8$ GeV. We also observe the
resonant enhancement of CP violation when $M_H=M_A$. More promising is
the reaction (b), in which CP violation may be observed for a wider
range of Higgs-boson masses, {\em i.e.}, for $M_H=350$ GeV $-$ 430
GeV.  Again, the mechanism of resonant CP violation is very important
to render the observable ${\cal A}_{CP}$ measurable, as shown in Fig.\ 
2.
  
Note that resonant CP-violating $HA$ transitions can also take place within
the MSSM, in which neutralinos and charginos may play the r\^ole of heavy
Majorana fermions. In the MSSM, the Higgs-mixing mass parameter $\mu$ in the
superpotential and the tri-linear soft-SUSY-breaking couplings $A$ can have
non-trivial CP-violating phases, which give rise to complex chargino- and
neutralino-mass matrices \cite{KO} and hence to interactions of the form given
in Eqs.\ (\ref{ANN}) and (\ref{HNN}). Yet, electric dipole moment
bounds on neutron and electron cannot prevent these CP-violating phases from 
being large \cite{KO}. Although multi-Higgs models, such as the Weinberg's
three-Higgs doublet model, may also induce a sizeable $HA$ mixing ({\em cf.}\
Eq.\ (\ref{HZmix})) for a certain corner of their parameter space, the
requirement of having resonant CP violation through a relatively small mass
difference between $H$ and $A$ will strongly favour only extended Higgs
sectors with a SUSY origin \cite{SUSYGUT,JFG}. Finally, we must stress that,
even though building a muon machine with a high degree of polarization may
become a difficult task, our analysis will, however, carry over to searches
for resonant CP-violating effects in the decay products of the final states,
{\em i.e.}, in the angular-momentum distributions and energy asymmetries of
the produced charged leptons and jets \cite{Peskin}.

\newpage

\centerline{\Large\bf Figure Captions }
\vspace{-0.2cm}
\newcounter{fig}
\begin{list}{\rm\bf Fig. \arabic{fig}: }{\usecounter{fig}
\labelwidth1.6cm \leftmargin2.5cm \labelsep0.4cm \itemsep0ex plus0.2ex }

\item CP-violating $HA$ and $HZ$ (for $A\equiv G^0$) transitions in
$\mu^+\mu^-$ collisions. 

\item Numerical estimates of production cross-sections and CP violation 
for $\mu^-_{L,R}\, \mu^+_{L,R}\to (H,\ A) \to f\bar{f}$ in two different 
SUSY scenarios with heavy Majorana neutrinos (see also text).

\end{list}

\end{document}